\documentclass[showkeys,showpacs,aps,amssymb,prd]{revtex4}

\usepackage{amssymb}
\usepackage{amsfonts}
\usepackage{amsbsy}
\usepackage{graphicx}
\usepackage{amssymb}
\usepackage{textcomp}
\usepackage{amsfonts}
\usepackage{graphicx}
\usepackage{amsmath}
\usepackage{amsthm}
\usepackage{amssymb}
\usepackage{ifthen}
\newcounter{multi} \newcounter{multa}

\newcounter{faki} \newcounter{faka}

    \newtheorem{theorem}{Theorem}
    
    \newtheorem{proposition}[theorem]{Proposition}
    \newtheorem{corollary}[theorem]{Corollary}

\theoremstyle{definition} 

\theoremstyle{remark} 

\newcommand{\be}{\begin{equation}}

\newcommand{\ee}{\end{equation}}

\newcommand{\beq}{\begin{eqnarray}}
\newcommand{\eeq}{\end{eqnarray}}

\newcommand{\beqa}{\begin{eqnarray}}
\newcommand{\eeqa}{\end{eqnarray}}

\newcommand{\beqn}{\begin{eqnarray*}}
\newcommand{\eeqn}{\end{eqnarray*}}

\newcommand{\beqan}{\begin{eqnarray*}}
\newcommand{\eeqan}{\end{eqnarray*}}

\newcommand{\comment}[1]{}

\newcommand{\R}{{\mathbb R}}

\newcommand{\N}{{\mathbb N}}

\newcommand{\area}{\operatorname{area}}

\newcommand{\RR}{{\mathbb R}}

\newcommand{\ra}{\rightarrow}

\def\tlda{{\tilde{a}}}

\def\tldh{\tilde{h}}
\def\tldk{\tilde{k}}

\def\tlddh{\tilde{\tilde{h}}}
\def\tlddk{\tilde{\tilde{k}}}

\begin{document}

\title{A Mathematical Theory of  Stochastic  Microlensing I.\\Random Time-Delay Functions and Lensing Maps}

\author{A. O. Petters}
\email{petters@math.duke.edu} \affiliation{Departments of
Mathematics and Physics, Duke University, Science Drive, Durham,
NC 27708, United States of America}

\author{B. Rider}
\email{brider@euclid.colorado.edu} \affiliation{    Department of
 Mathematics,
    University of Colorado at Boulder,
    Campus Box 395
    Boulder, CO 80309,
    United States of America}

\author{A. M.
Teguia}\email{alberto@math.duke.edu} \affiliation{Department of
Mathematics, Duke University, Science Drive, Durham, NC 27708,
United States of America}


\begin{abstract}
Stochastic microlensing is a central tool in probing dark matter
on galactic scales. From first principles, we initiate the
development of  a mathematical theory of stochastic microlensing.
Beginning with the  random time delay function and associated
lensing map, we determine exact expressions for the  mean and
variance of these transformations. In addition, we derive the
probability density function (p.d.f.) of a random point-mass
potential, which form the constituent  of a stochastic microlens
potential.  We characterize the exact p.d.f. of a normalized
random time delay function at the origin, showing that it is a
shifted gamma distribution, which also holds  at leading order
 in the limit of a large number of point masses if  the
normalized time delay function was at a general point of the lens
plane. For the large number of point masses limit, we also prove
that the asymptotic p.d.f. of the random lensing map under a
specified scaling converges to a bivariate normal distribution. We
show analytically that the p.d.f. of the random scaled lensing map
at leading order depends on the magnitude of the scaled bending
angle due purely to point masses as well as demonstrate explicitly
how this radial symmetry is broken at the next order.
Interestingly, we found at leading order a formula linking the
expectation and variance of the  normalized random time delay
function to the first Betti number of its domain. We also
determine an asymptotic p.d.f. for the  random bending angle
vector and find an integral expression for the probability of a
lens plane point being near a fixed point. Lastly, we show
explicitly how the results are affected by location in the lens
plane. The results of this paper are relevant to the theory of
random fields and provide a platform for further generalizations
as well as analytical limits for checking astrophysical studies of
stochastic microlensing.
 \end{abstract}



\maketitle

\section{Introduction}
Stochastic microlensing is the study of the deflection of light by
a random collection of stars. In recent years, this subject has
become  a central tool in understanding the nature and
distribution of dark matter on galactic scales (e.g.,  Schechter
and Wambsganss \cite{Schechter1}).  This has raised the need for a
rigorous framework for stochastic microlensing.

Research on stochastic microlensing has been mainly numerical, or
at least semi-analytical, in flavor. Work in the early 1980's
included an analytical study of certain stochastic effects  of
microlensing by a single random lens  (e.g., Vietri and Ostriker
\cite{Vietri}). This was later extended to the case of multiple
lenses  using numerical simulations (e.g., Kayser, Refsdal, and
Stabell \cite{Kayser}; Paczynski \cite{Pac}).  While successful in
furthering our picture of stochastic microlensing, these
simulations often cannot be checked analytically and are
time-consuming. Some of the first analytical counterparts of the
numerics were focused  on the study of  random surface
brightnesses, bending angles, magnifications, and redshifts (e.g.,
Deguchi and Watson \cite{Deguchi,Deguchi2}; Katz, Balbus, and
Paczynski  \cite{Katz};  Schneider \cite{Schneider}; Turner,
Ostriker, and Gott \cite{Turner};  Vietri \cite{Vietri1}). This
was followed by numerical and semi-rigorous work characterizing
the  magnification probabilities and the expected number of
microimages in a variety of  lensing situations (e.g., Rauch {\it
et al.} \cite{Rauch}; Wambsganss \cite{Wambsganss}; Wambsganss,
Witt, and Schneider \cite{wamb}). Recently, similar techniques
were employed to study the distributions of microimages,
magnifications, and  time delay differences between images, with
the last two   applied directly to probe dark matter using
flux-ratio anomalies (e.g., Granot, Schechter, and Wambsganss
\cite{GSW};  Keeton, and Moustakas \cite{Keeton}; Schechter, and
Wambsganss \cite{Schechter1};  Schechter, Wambsganss, and Lewis
\cite{Schechter2}; Tuntsov {\it et al.} \cite{Tuntsov}). In a more
general context,  Berry and collaborators also explored the
statistical aspects of caustics in optical systems for the case of
Gaussian random fields (see \cite{Berry} and references therein).

This series  aims to  develop a mathematical theory of stochastic microlensing,
starting from first principles. This fits in the general
theory of random fields, namely, the study of random
functions and  their critical point structure (for more on random fields, see Adler and Taylor
\cite{Adlerjon} and references therein). We derive new stochastic results,
for a general point in the lens plane,  about
random time delay functions and lensing maps, which are transformations
forming the core of microlensing theory. We determine the exact
(as opposed to asymptotic) expectation and variance of
 both the lensing map and the point mass potential. Using the expected value of the
 components of the  lensing map, we deduce, for a fixed image in the lens plane,  the
  corresponding average position of the point it is mapped to in the light source plane.
  Furthermore, since  point mass potentials form the density components of the  time delay
 function, we use the  exact formulas of their first and second moments to derive the asymptotic
expectation and variance of the normalized time delay function in the large number of stars limit.
We also reveal  an interesting topological link, namely, at
leading order we show that the expectation and variance of the normalized random microlensing time delay function
is related to  the first Betti number of the time delay functions's domain.

Continuing our study of the normalized random  time delay function, we establish that at the origin its exact p.d.f.
is a shifted gamma distribution, and  that this is asymptotically the case for an
 arbitrary  point within the lens plane  in the large number of stars limit.
  In the same limit, we prove that the p.d.f of a scaled lensing map converges to a
bivariate Gaussian, with once again consistent
 asymptotic mean and  variance. This analytical work
shows  that the p.d.f. of the random scaled lensing map at leading order  term depends on the square of  magnitude of the
scaled bending angle due purely to point masses. It also illustrate how  this radial symmetry is broken at the next order.
Similarly, we note that the p.d.f. of the random time delay function at the origin
depends only on the potential time delay due purely to point masses, a result also holding at leading order
for an arbitrary point in the lens plane.

 Taking our analysis further, we compute the probability that the image of  a given point
 under the lensing map will be  in a given disk  in light the
 source plane, which provides information about the light source position.  We deduce from
 these results the expectation, variance, and asymptotic p.d.f. of the bending angle vector
at an arbitrary position in the lens plane. We also use these
results to compute the probability that a  point in the lens plane
will be mapped in a certain neighborhood of ``itself'' in the
light source plane. This study relates to the theory of fixed
points, which was studied earlier by  Petters and Wicklin
\cite{Fixe} in the case of a deterministic lens system.

Overall, these first steps lay the foundation in terms of concepts, results, and tools
for future analytical work on random time delay functions and lensing mappings,
including lensing observables like magnification,
arising from more complex stochastic mass distributions.
The work can also help as a guide on tractable aspects of the general theory
of random mappings.

We  state the basic random microlensing framework in Section~\ref{basic}. In
Section~\ref{time}, we present the results of our study of the   random time delay function,
obtaining its expectation, variance, exact p.d.f. at the origin, and asymptotic p.d.f.
at an arbitrary lens plane point.  A
similar presentation of results is done  for  the random lensing map   in
Section~\ref{lens}. In Section~\ref{angle}, we apply our results
about random lensing maps to random bending angle vectors and fixed points.
The detailed proofs of these results are given in the Appendix.

Finally, the second paper in this program will explore the random shear and
 expected number of microimages produced by stochastic microlensing.

\section{Basic framework}\label{basic}

Consider  a microlensing situation where the light source is
point-like and located at $y$ in $\RR^2$ (the light source plane)
and  the gravitational lens is a collection of $g$
point masses (stars) that are  randomly
distributed over a region of the
lens plane.
This lensing scenario
is modeled with the following gravitational lens potential, where
all quantities are in dimensionless form (see \cite[pp. 79, 104]{Petters}):

\beq
\label{defP} \psi_g(x)=\frac{\kappa_c}{2} |
 x |^2 -\frac{\gamma}{2}(x_1^2-x_2^2)
       + \sum_{j=1}^g m_j  \log | x - \xi_j |,
\eeq
with $x = (x_1, x_2)$  in the lens plane $L= \{\xi_1, \dots, \xi_g\}$.

In a {\it standard stochastic microlensing} situation, the light
source position $y$ is assumed to be a uniform random vector in
$\RR^2$ and the star positions $\xi_j=(U_j,V_j)$ are taken to be
independent, identically distributed (i.i.d.)  uniform random
vectors in  a region of the lens plane.  The continuous
matter density $\kappa_c$, external shear $\gamma$, masses $m_j$,
and number of stars $g$ are assumed fixed, unless stated to the
contrary. The random lens potential (\ref{defP}) is widely used in
studies of stochastic lensing due to stars (see
\cite{SchEhlFal92,Petters}).

\smallskip

\noindent {\bf Remark:}
\vspace{-0.1in}
\begin{itemize}
\item The constants $m_j$ are  scalar multiples of the actual
stars' physical masses, but we shall continue to call them masses
 for simplicity (see \cite[p. 102]{Petters}).
\item A natural next step is to consider a lensing scenario  where the masses are
distributed isothermally in an elliptical galaxy. This is a work in progress.
\end{itemize}

Gravitational microlensing of a light  source  at position $y$
in $\RR^2$ causes  a change in a
 light ray's arrival time that is captured analytically by the dimensionless
random time delay function given  as follows (see \cite{Petters},
p. 81):
\be
\label{defT}
     T_{g,y}(x) =   \frac{1}{2} | x - y |^2 -\frac{\kappa_c}{2} |
 x |^2 +\frac{\gamma}{2}(x_1^2-x_2^2)
       - \sum_{j=1}^g m_j  \log | x - \xi_j |.
\ee

\noindent \textit{Fermat's Principle of stationary  time} states
(see \cite{Petters}, p. 67): \textit{Light rays from the source at
$y$ to the observer are characterized by critical points of the
time delay function $T_y(x)$}, that is, the solutions $x$ in the lens plane $L$,
where $L = \RR^2 - \{\xi_1, \dots, \xi_g\}$,
of the equation

\beq
\label{Fermat}
\nabla T_y(x)= {\bf 0},
\eeq
where the gradient $\nabla$ is with respect to the $x$ variable.

\noindent The  corresponding dimensionless random lensing map
$\eta_g :L\ra\R^2$,
(the lens plane),
 is defined  by (see \cite{Petters}, p. 81):
\[\eta_g(x)=x-\nabla\psi_g(x).\]

\noindent The advantage  of the lensing map is that solutions $x$
of (\ref{Fermat}) are equivalent to solutions $x$  of the {\it
lens equation} (see \cite{Petters}, p. 81):
$$\eta_g(x)=y,$$
\noindent which are preimages $x$ of $y$ under the lensing  map
$\eta_g$. Solutions $x\in L$ of the lens equation are called {\it
lensed images}.

Microlensing also causes light rays to bend. For a light ray
passing through the point $x$ in the lens plane, the {\it bending
angle} $\alpha_g(x)$  at $x$ is given by (see \cite{Petters},
p. 79):
\[
\alpha_g(x)=\nabla\psi_g(x).
\]

\subsubsection*{\bf Assumptions and Notation}

We shall abide by  the following throughout the paper:
\begin{itemize}

\item Rectangular coordinates $(u,v)$ are assumed in  the lens
plane $L$.

\item All masses are  equal, $m_j\equiv m\neq 0,$ for $
j=1,\cdots,g$.

\item The star positions $ \xi_i = (U_i, V_i)$ are independent and
uniformly distributed in  the disk $B({\bf 0},R)$ of radius $ R$
centered at ${\bf 0}$, that is,
\[
     (U_i, V_i)  \sim  {\rm Unif} \Bigl\{ (u,v)\in L :  |u|^2 + |v|^2 \le R^2
 \Bigr\},  \  \  \  \   \ i = 1, \cdots, g.
\]

\item The quantities $R$ and $g$ are related by the following
physical formula for  {\it the surface mass density} $\kappa_*$:

\beq
\label{kappa}
\kappa_*&=& \frac{m~ g}{R^2}.
\eeq
Unless
otherwise stated assume that $\kappa_*$ is fixed.

\item {\it Notation}: For clarity, the  statement
$$
\mbox{ ``} F(x) = G(x) + O( K(x)) \quad \mbox{as} \ x \ra \infty
\mbox{''}
$$
means precisely that
$$
\lim_{x \ra \infty} \frac{|F(x) - G(x)|}{K(x)} \le C
$$
for some constant $C >0$.

\item {\it Notation:}  $\log^\ell A = (\log A)^\ell.$

\end{itemize}

\section{Results on The Random time delay function}\label{time}

We can  write  the time delay function  as
\[
T_{g,y}(x) =   \frac{1}{2} | x - y |^2\ -\ \frac{\kappa_c}{2} | x |^2\    + \ \frac{\gamma}{2}(x_1^2-x_2^2)\ + \ \sum_{j=1}^g W_{g,j}(x),
\]
       where
\[
W_{g,j}(x)=-m  \log | x - \xi_j |
\]
is the point mass potential.

We  define a  ``normalization" $T_{g,y}^{*}$ of $T_{g,y}$ as
follows:
\[
T_{g,y}^{*}(x)\equiv T_{g,y}(x)+ g\, m\log R=\frac{1}{2} | x - y |^2
-\frac{\kappa_c}{2} | x |^2 +\frac{\gamma}{2}(x_1^2-x_2^2)\ + \ \sum_{j=1}^g W_{g,j}^{*}(x),
\]
where
\[
W_{g,j}^{*}(x)=-m  \log \frac{| x - \xi_j |}{R}
\]
is the {\it normalized  point mass potential}.  Note that this
normalization (of the time delay function) does not  change the lensing map and relevant
physical lensing quantities such as magnification, difference in
arrival times between two images, etc.

\subsection{Exact Moments and Probability Density Functions }

The  first and second moments of a random point mass potential are
given by:

\begin{proposition}
\label{propp}
 Let $x\in B({\bf 0},R)$ and $y\in \R^2$ \mbox{\rm(}the light source plane\mbox{\rm)}. Then
\[
E[W_{g,j}(x)]=\frac{\kappa_*(R-|x|)^2(1-2\log (R-|x|))}{2g}
-\frac{2\kappa_*}{\pi g }\int_{R-|x|}^{R+|x|} r f(r,|x|)(\log r)
dr,
\]
 and

\beqn
E[W_{g,j}^2(x)]&=&\frac{m\kappa_*(R-|x|)^2 \left[1+2(\log (R-|x|)-1)\log(R-|x|)\right]}{2 g} + \frac{2m\kappa_*}{\pi g}\int_{R-|x|}^{R+|x|} r f(r,|x|) (\log^2 r)  dr,
\eeqn
where
$f(r,|x|)={\rm Arccos}\Bigl(\frac{|x|^2+r^2-R^2}{2|x|r}\Bigr)$ for
non-zero $x$, and $f(r,0)=0$.
\end{proposition}
\vspace{0.3in}
\begin{proof}[Proof:]
See Appendix.
\end{proof}
\vspace{0.3in}

Both integrals in the above proposition are finite (and can be
computed  numerically). Hence, the expectation  of $T_{g,y}(x)$,
\[
E[T_{g,y}(x)]=\frac{1}{2} | x - y |^2-\frac{\kappa_c}{2} | x |^2 +\frac{\gamma}{2}(x_1^2-x_2^2)+gE[W_{g,j}(x)]
\]
and
its variance
\[
{\rm Var}[T_{g,y}(x)]=gE[W_{g,j}^2(x)]-g(E[W_{g,j}(x)])^2
\]
are also finite. Instead of pursuing  exact expressions for the
mean and the variance of  $T_{g,y}^*(x)$ for general $x\in B\left({\bf 0},R\right)$, we take an asymptotic
approach:

\begin{corollary}
\label{check1}
Let $x\in B({\bf 0},R)$ and $y\in \R^2$ \mbox{\rm(}the light
source plane\mbox{\rm)}. In the large $g$ limit, we have
\[
E[T_{g,y}^*(x)]= \frac{m}{2}\, g \ + \ O(g^{1/2}\log g)
\]
and
\[
{\rm Var}[T_{g,y}^*(x)]= \left(\frac{m}{2}\right)^2 g \ + \ O(g^{1/2}\log^2 g).
\]
\end{corollary}
\vspace{0.3in}

\begin{proof}[Proof:]
 See Appendix.
\end{proof}
\vspace{0.3in}

Corollary~\ref{check1} shows an interesting link that has a ``stochastic Morse theoretic'' flavor, namely, at leading order,
$$
E[T_{g,y}^*(x)]= \frac{m}{2} B_1(D_*), \qquad
{\rm Var}[T_{g,y}^*(x)]= \left(\frac{m}{2}\right)^2 B_1(D_*)
$$
where $D_*=B\left({\bf 0},R\right)-\{\xi_1,\cdots,\xi_g\}$ and $B_1(D_*)$ is the first Betti number of $D_*$. This is because
 $B_1(D_*)=g$, which is the number of ``holes'' in $D_*$ (e.g., \cite[p. 399]{Petters}).

We now derive the p.d.f.'s of the point mass potentials $W_{g,j}(x)$ and
$W_{g,j}^*(x)$:

\begin{proposition}
\label{expo}
For $x\in B({\bf 0},R)$, the p.d.f. of $W_{g,j}(x)$
 is given by:

\beq
\label{pdf-pmpot0}
f_{W_{g,j}(x)}(h)&=&\left\{
\begin{array}{lr}
\frac{2}{mR^2}\exp[-\frac{2h}{m}],&~~~~~-m\log (R-|x|)<h\\
A_{R,x}^{'}(h),&~~~~-m\log (R+|x|)<h< -m\log (R-|x|)\\
0,&~~~~h< -m\log (R+|x|),
\end{array}
\right.
\eeq
 where

\beqn
A_{R,x}(h)  &=&1-\Bigl(r_h\cos^{-1}\theta_{R,1}(h)+R^2\cos^{-1}\theta_{R,2}(h)-B_{R,x}(h)\Bigr)/(\pi R^2),\\
 r_h &=&\exp[2h/m],\\
\theta_{R,1}(h)&=& (|x|^2+r_h^2-R^2)/(2 |x| r_h),\\
\theta_{R,2}(h)&=&(|x|^2+R^2-r_h^2)/(2 |x| R),\\
B_{R,x}(h)&=&\sqrt{(R+r_h-|x|)(d+r_h-R)(d+R-r_h)(d+r_h+R)}/2.
\eeqn
It then follows that the p.d.f. of  $W_{g,j}^*(x)$ is:

\beq
\label{pdf-pmpot}
f_{W_{g,j}^*(x)}(h)=\left\{
\begin{array}{lr}
\frac{2}{m}\exp[-\frac{2h}{m}],&~~~~~-m\log (1-\frac{|x|}{R})<h\\
A_{1,\frac{x}{R}}^{'}(h),&~~~~-m\log (1+\frac{|x|}{R})<h< -m\log (1-\frac{|x|}{R})\\
0,&~~~~h< -m\log (1+\frac{|x|}{R}).
\end{array}
\right.
\eeq
\end{proposition}

\vspace{0.3in}

\begin{proof}[Proof:]
 See Appendix.
\end{proof}
\vspace{0.3in}

 At the origin, we obtain the exact p.d.f. for the random time delay function:

\begin{corollary}
\label{cor:pdf-TDstar0}
 The random variable $T_{g,y}^{*}({\bf
0})$ has a  shifted gamma  distribution with density

\beq
\label{pdf-TD0}
f_{T_{g,y}^{*}({\bf 0})}(h)&=&\left\{
\begin{array}{lr}
\Bigl(\frac{2}{m}\Bigr)^g \frac{(h-c)^{g-1}}{(g-1)!}\exp\left[-\frac{2(h-c)}{m}\right],&~~~~~h>c\\
~~0 ,&~~~~~h< c,
\end{array}
\right.
\eeq
where $c=|y|^2/2$.
\end{corollary}
\vspace{0.3in}

\begin{proof}[Proof:] From
Proposition \ref{expo}, we know that $W_{g,j}^*(0)$  is an
exponential random variable with parameter $2/m$. By independence,
$\sum_{j=1}^g W_{g,j}^{*}(0)\sim \Gamma (g,2/m).$
\end{proof}
\vspace{0.3in}

A direct calculation using the result of
Corollary~\ref{cor:pdf-TDstar0} shows that the exact expectation
and variance of the   normalized random time delay function
at the origin are given respectively as follows:
\[
E[T_{g,y}^*({\bf 0})]= \frac{m}{2}\, g\ + \ c,
\qquad
{\rm Var}[T_{g,y}^*({\bf 0})]= \left(\frac{m}{2}\right)^2 g,
\]
which is consistent with Corollary~\ref{check1}.

Although  Corollary~\ref{cor:pdf-TDstar0} only gives the p.d.f. of $T_{g,y}^{*}({\bf
0})$, it will play a central role in establishing the asymptotic
p.d.f. of  $T_{g,y}^{*}(x)$ for an arbitrary $x\in B({\bf 0},R)$
--- see Theorem~\ref{thm:pdf-TDstar} below.

\subsection{Asymptotic P.D.F. for the Time Delay Function}

The expression of the p.d.f. $f_{T_{g,y}^{*}(x)}$ for arbitrary
$x$ can be derived via  convolution and equation~(\ref{pdf-pmpot}), but is very cumbersome. As $g\ra \infty$ though,
the analysis simplifies.

\begin{theorem}
\label{thm:pdf-TDstar}
For every $x\in B({\bf 0},R)$,  let
$f_{T_{g,y}^{*}(x)}$ be the p.d.f. of
$T_{g,y}^{*}(x)$. In the large $g$ limit, we have:

\beq
\label{pdf-TDstar}
f_{T_{g,y}^{*}(x)}(h)&=&f_{T_{g,y}^{*}({\bf
0})}(h-d_1(x,y))+O(g^{-3/2})\nonumber\\
&=&\left\{
\begin{array}{lr}
\Bigl(\frac{2}{m}\Bigr)^g \frac{(h-d_1-c)^{g-1}}{(g-1)!}\exp\left[-\frac{2(h-d_1-c)}{m}\right],&~~~~~h>d_1+c\\
~~0 ,&~~~~~h< d_1+c,
\end{array}
\right. \ \ \ \ \ \ O(g^{-3/2})
\eeq
where $d_1(x,y)= \frac{1}{2} | x - y |^2-\frac{\kappa_c}{2} | x |^2 +\frac{\gamma}{2}(x_1^2-x_2^2)$.
\end{theorem}

\vspace{0.3in}

\begin{proof}[Proof:]
 See Appendix.
\end{proof}
\vspace{0.3in}

Theorem~\ref{thm:pdf-TDstar} shows that, at leading order, the p.d.f.
$f_{T_{g,y}^{*}(x)}$ of the normalized time delay function
$T_{g,y}^{*}(x)$ is a gamma distribution. Moreover, since
$$
h-d_1(x,y)-c=\sum_{j=1}^g-m\log\left(\frac{|x - \xi_j|}{R}\right),
$$
the p.d.f. $f_{T_{g,y}^{*}(x)}$ for any $x\in B({\bf 0},R)$
depends at leading order on the  normalized potential time
delay due purely to point masses.

 \section{ Results on the Random lensing map}\label{lens}

The lensing map $\eta_g :L\ra\R^2$ is defined by
\[
\eta_g(x)=x-\nabla\psi_g(x),
\]
where $L = \RR^2 - \{\xi_1, \dots, \xi_g\}$ is the lens plane. The
components of the lensing map are given by
\[
     \eta_{1,g}(x)  = (1-\kappa_c+\gamma)x_1 + \sum_{j=1}^g  \frac{m(U_j -
 x_1)}{R_j^2(x)}, \qquad \eta_{2,g}(x) = (1-\kappa_c-\gamma)x_2  +
\sum_{j=1}^g
  \frac{m(V_j - x_2)}{R_j^2(x)}
\]
where $R_j^2(x) = (U_j - x_1)^2 + (V_j - x_2)^2$.

\subsection{Exact Expectation and Variance of the Lensing Map Components}

The components $\eta_{1,g}(x)$ and $\eta_{2,g}(x)$ of the lensing
map have the following basic statistics:

\begin{proposition}
\label{meaneta}
 For $x=(x_1,x_2)\in L\cap B({\bf 0},R)$, the exact expectations of the components of
the lensing map are
\[
E[\eta_{1,g}(x)]= \left[1- (\kappa_c + \kappa_*) +\gamma\right]\, x_1
~~~~~~~{\rm and}~~~~~~
E[\eta_{2,g}(x)]= \left[1- (\kappa_c +\kappa_*) -\gamma\right]\, x_2 .
\]
Both variables have infinite variance.
\end{proposition}

\vspace{0.3in}

\begin{proof}[Proof:]
 See Appendix.
\end{proof}
\vspace{0.3in}

Proposition~\ref{meaneta} provides statistical information about
the position values of the lensing map in the light source plane.
In fact, suppose that an image position $x=(x_1,x_2)$ is given.
The proposition gives:
\[
E[\eta_{1,g} (x_1)]=  \left[1-(\kappa_c + \kappa_*) +\gamma\right]\, x_1,
 \qquad
E[\eta_{1,g} (x_2)] = \left[ 1- (\kappa_c + \kappa_*)
-\gamma\right]\, x_2 .
\]
If $x = {\bf 0}$, then on average the
values of $\eta_g ({\bf 0})$ are isotropically distributed around
the origin in the light source plane.  More generally, we obtain
the following:

\begin{itemize}
\item {\it Case I} (``macro-minimum''): $1-(\kappa_c + \kappa_*)
+\gamma > 0$ and $1-(\kappa_c + \kappa_*) - \gamma > 0$. If $x$
lies in a given quadrant of the lens plane, then on average the
random lensing map will send $x$ over to the same quadrant in the
light source plane.

\item {\it Case II} (``macro-saddle''): $1-(\kappa_c + \kappa_*)
+\gamma > 0$ and $1-(\kappa_c + \kappa_*) - \gamma <  0$. If $x$
lies in quadrant I of the lens plane, then on average the random
lensing map will send $x$ over to quadrant IV in the light source
plane. Similarly, we obtain results for the other quadrants.

\item {\it Case III} (``macro-maximum''): $1-(\kappa_c + \kappa_*)
+\gamma < 0$ and $1-(\kappa_c + \kappa_*) - \gamma < 0$. If $x$
lies in given quadrant of the lens plane, then on average the
random lensing map will send $x$ over in the light source plane to
the quadrant symmetric with respect to the origin (e..g, quadrant
I goes to III on average).

\end{itemize}

\subsection{Asymptotic P.D.F. of the Lensing Map}

For the remainder of the paper, we shall  assume that
$$ R^2 = g/\pi, \qquad g >1,$$
which gives $ \kappa_*=\pi m$ and ignores the known case with $g =
1$ (single star lens).  Also, we suppose that the random lensing
map has an  absolutely continuous cumulative distribution
function.  Then:

\begin{theorem}
\label{thm-pdf-lm}
Let $x=(x_1,x_2) \in L\cap B\left( {\bf 0},R\right)$ be fixed. The p.d.f. of
the scaled lensing map $\frac{\eta_g(x)}{\sqrt{\log g}}$ is given
in the large $g$ limit by:

\beq
\label{resultthm-pdf-lm11}
f_{\frac{\eta_g(x)}{\sqrt{\log g}}}(\tldh, \tldk)
&=&\frac{e^{-\frac{(\tldh-\tilde{a}_1)^2+ (\tldk-\tilde{a}_2)^2}
{2 \tilde{\sigma}_g^2}}}{(\sqrt{2\pi}\tilde{\sigma}_g)^2}
\left[ 1 \ - \ \frac{  \kappa_*}{\sqrt{\log g}}\
\frac{x_1(\tldh-\tilde{a}_1)+x_2(\tldk-\tilde{a}_2)}{\tilde{\sigma}_g^2}\right.
~~~~~~~~~~~~~~~\nonumber\\
&&\left. ~~~~~~~~~~~~~~~+ \frac{\kappa_*^2}{4 \pi} \
\frac{((\tldh-\tilde{a}_1)^2+(\tldk-\tilde{a}_2)^2)-
2\tilde{\sigma}_g^2}{\tilde{\sigma}_g^4}\ \frac{\log(\log g)}{\log g}\right]
+ O\Bigl(\frac{1}{\log g}\Bigr),
\eeq

 where $(\tldh, \tldk)$ are the possible values of $\frac{\eta_g(x)}{\sqrt{\log g}}$ and
\beqan
\tilde{a}_1 =\frac{(1-\kappa_c+\gamma)x_1}{\sqrt{\log g}}, & & \qquad
\tilde{a}_2 = \frac{(1-\kappa_c-\gamma)x_2}{\sqrt{\log g}},
\\
&& \\
\tilde{\sigma}_g=\frac{\sigma_g}{\sqrt{\log g}}, && \qquad
\sigma_g = \frac{\kappa_*}{\sqrt{\pi}} \, \sqrt{\log(B\,
g^{1/2})}, \qquad
 B =  \frac{2\sqrt{\pi}e^{1-\gamma_e}}{\kappa_*}\, (\gamma_e \ \mbox{is the
Euler constant}).
\eeqan
The  p.d.f.s of the components of the scaled lensing map are:
\[
f_{\frac{\eta_{i,g}(x)}{\sqrt{\log g}}}(\tilde{c})
=\frac{e^{-\frac{(\tilde{c}-\tilde{a}_i)^2}{2\tilde{\sigma}_g^2}}}{\sqrt{2\pi}\,\tilde{\sigma}_g}
\Bigl[1-\frac{\kappa_*}{\sqrt{\log g}}\frac{x_i (\tilde{c}-\tilde{a}_i)}{\tilde{\sigma}_g^2}+
\frac{\kappa_*^2}{4 \pi}\frac{(\tilde{c}-\tilde{a}_i)^2-\tilde{\sigma}_g^2}{\tilde{\sigma}_g^4}
\,\frac{\log(\log g)}{\log g}\Bigr]+ O\Bigl(\frac{1}{\log g}\Bigr)
\]
for $i=1,2.$   For notational simplicity, we use $\tilde{c}$ to represent the possible values of both
 $\eta_{1,g}(x)$ and $\eta_{2,g}(x)$.
\end{theorem}

\vspace{0.3in}
\begin{proof}[Proof:]
 See Appendix.
\end{proof}
\vspace{0.3in}

Observe that the p.d.f. of the scaled lensing map
$f_{\frac{\eta_g(x)}{\sqrt{\log g}}}$ depends on $x$ in its leading order term.
Furthermore, the fact that
$$
{\rm Var}\left[\frac{\eta_{i,g}(x)}{\sqrt{\log
g}}\right]=\infty, \qquad i=1,2,
$$
makes applying a standard
Central Limit Theorem unwieldy, thus the method of proof presented in the Appendix.

Theorem~\ref{thm-pdf-lm} shows that at leading order, the p.d.f. of the scaled
lensing map is bivariate normal and illustrates how at the next order the p.d.f. deviates from normality. Moreover, since
$$
(\tldh-\tlda_1, \tldk-\tlda_2) = \frac{\alpha_g(x)}{\sqrt{\log g}} \equiv \tilde{\alpha}_g^*,
$$
which is the scaled bending angle due purely to stars, the quantity
$(\tldh-\tlda_1)^2+(\tldk-\tlda_2)^2$
in (\ref{resultthm-pdf-lm11}) is $|\tilde{\alpha}_g^*|^2$.  In other words,
the leading factor of the p.d.f. of the scaled lensing map depends on the square of the magnification of the
scale bending angle due to stars.  This radial symmetry is broken at the next order term
since there is a dependence on the individual components
of the scaled bending angle.

We now discuss several consequences of Theorem~\ref{thm-pdf-lm}.

\begin{corollary}
\label{corthm-pdf-lm}
As $g \ra \infty$ with fixed $x \in \R^2$, we have:
\[
     \Bigl( \frac{ \eta_{1,g}(x)}{ \sqrt{\log g} }   ,  \frac{\eta_{2,g}(x)}{\sqrt{\log
 g}}   \Bigr)
      \Rightarrow   ( \eta_{1,\infty}, \eta_{2,\infty} ).
\]
Here ``$\Rightarrow$'' denotes distributional convergence and
$(\eta_{1,\infty}, \eta_{2,\infty} )$ is a bivariant normal random
vector with independent mean-zero normal random variables as
components, each with variance $\kappa_{*}^2/(2\pi)$.
\end{corollary}

\vspace{0.3in}

\begin{proof}[Proof:]
 See Appendix.
\end{proof}
\vspace{0.3in}

We can readily obtain the p.d.f. of the
unscaled lensing map:

\begin{corollary}
\label{thm-pdf-lm11}
Let $x=(x_1,x_2) \in L$ be fixed. The p.d.f.
of $\eta_g(x)$ is given in the large $g$ limit by:

\beq
\label{resultthm-pdf-lm}
f_{\eta_g(x)}(h,k) &=&\frac{e^{-\frac{(h-a_1)^2+ (k-a_2)^2}
 {2\sigma_g^2}}}{\Bigl(\sqrt{2\pi}\sigma_g\Bigr)^2}
\Bigl[1-\kappa_*\frac{x_1(h-a_1)+x_2(k-a_2)} {\sigma_g^2} \nonumber\\
&& ~~~~~~~~+ \frac{\kappa_*^2}{4 \pi}\
\frac{\Bigl((h-a_1)^2+(k-a_2)^2-2\sigma_g^2\Bigr)}{\sigma_g^4}H(g)\Bigr]
+ O\Bigl(\frac{1}{\log^2 g}\Bigr),
\eeq
with the possible values
of the random vector $\eta_g(x)$ written as $(h,k)$.  Here
$\sigma_g$ and $H(g)$ are as defined in Theorem~\ref{thm-pdf-lm},
and
$$
a_1 =(1-\kappa_c+\gamma)x_1, \qquad a_2 =(1-\kappa_c-\gamma)x_2,
\qquad H(g) = \log(\log g).
$$
\end{corollary}

\vspace{0.3in}

\begin{proof}[Proof:]
First, note that the possible values of $\eta_g(x)$ and
$\frac{\eta_g(x)}{\sqrt{\log g}}$ are related by

\beqan
\tldh =\frac{h}{\sqrt{\log g}}, & & \qquad \tldk =\frac{k}{\sqrt{\log g}}.
\eeqan
We know that
$$
f_{\eta_g(x)}(h,k)~=~\frac{1}{\log g} f_{\frac{\eta_g(x)}{\sqrt{\log g}}}(\tldh,\tldk)
~=~\frac{1}{\log g} f_{\frac{\eta_g(x)}{\sqrt{\log g}}}
\left(\frac{h}{\sqrt{\log g}},\frac{k}{\sqrt{\log g}}\right).
$$
The result then follows directly from
 equation~(\ref{resultthm-pdf-lm11}).
\end{proof}
\vspace{0.3in}

\begin{figure}[htp]
\centering
  \includegraphics[width=3in]{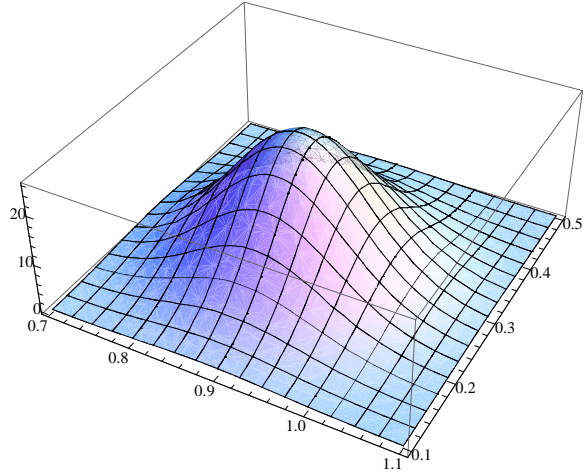}
\caption{ Asymptotic p.d.f. of the random lensing map in the large
number of stars limit.}
\label{fig:sub} 
\end{figure}

Let us consider the leading function in equation~(\ref{resultthm-pdf-lm}), namely,

$$
{\cal F}_{g,x}(h,k) = \frac{e^{-\frac{(h-a_1)^2+ (k-a_2)^2}{2\sigma_g^2}}}
{\Bigl(\sqrt{2\pi}\sigma_g\Bigr)^2}
\left[1-\kappa_*\frac{x_1(h-a_1)+x_2(k-a_2)} {\sigma_g^2} \ + \
\frac{\kappa_*^2}{4 \pi}\
\frac{\Bigl((h-a_1)^2+(k-a_2)^2-2\sigma_g^2\Bigr)}{\sigma_g^4}H(g)\right],
$$

\noindent where $x\in B({\bf 0},R)$ and $(h,k)\in \R^2$.  Note
that $H(g)>0$ if $g \ge 3$.  The function ${\cal F}_{g,x}$ is
actually a p.d.f. for $|x|$ sufficiently small and $g\geq 3$. In
fact, under the latter, we have ${\cal F}_{g,x}(h,k)\geq 0~$ for
all $(h,k)\in \R^2$. Moreover, integration-by-parts gives:
$$
\int_{\R^2} {\cal F}_{g,x}(h,k)\, dh dk = 1.
$$
Hence, the function ${\cal F}_{g,x}$ is a p.d.f. on $\R^2$ for
$|x|$ sufficiently small and $g\geq 3$.
Figure \ref{fig:sub}  depicts the graph of
${\cal F}_{g,x}$ with $x=(x_1,x_2)= (0.2,0)$,
$\kappa_c=0.405,~\gamma=0.3,~\kappa_*=0.045,$ and $g=10^6$.

Furthermore, there is enough control in the p.d.f. $f_{\eta_g(x)}$
to perform the expectation integrals to
obtain:

\begin{corollary} \label{cor:asym-lm-EV} Let $x\in L$ the lens plane. In  limit $g\ra\infty$, the expectations of the components $\eta_{1,g}(x),\eta_{2,g}(x)$ of the lensing map  are given by
 \[
E[\eta_{i,g}(x)]\simeq \left[1- (\kappa_c + \kappa_*) + (-1)^{i+1}
\gamma \right]\, x_i,  ~~~~i=1,2.
\]
Also, the variances of the lensing map components are infinite:
\[
{\rm Var}[\eta_{i,g}(x)]\simeq
\frac{\kappa_*^2}{\pi}[\log(B\,g^{1/2})+\frac{1}{2}\log\log g -\pi x_i^2]
= \sigma_g^2\ +\ \frac{\kappa_*^2}{2\pi}\log\log g \ -\  \kappa_*^2\,  x_i^2,\ \ g\ra\infty,
~~~~~i=1,2.
\]
\end{corollary}

\vspace{0.3in}

\begin{proof}[Proof:]
This result is obtained from Corollary~\ref{thm-pdf-lm11} (see equation~(\ref{resultthm-pdf-lm})) by computing the expected value and the variance of the random vector with density ${\cal F}_{g,x}(h,k)$.
\end{proof}
\vspace{0.3in}

Corollary~\ref{cor:asym-lm-EV} shows that the expectation
from the asymptotic p.d.f. ${\cal F}_{g,x}$ is consistent with the
findings in Proposition~\ref{meaneta}. We also have consistency  for the variance as $g\,\ra\,\infty$. Note that in
the leading order of Corollary~\ref{cor:asym-lm-EV}
the divergent variance depends on the position $x$.

Next, observe that given $g > 1$, the normal leading factor of
the p.d.f. $f_{\eta_g(x)}$ is maximized at
$$
{\bf a}=(a_1,a_2)=((1-\kappa_c+\gamma)x_1,(1-\kappa_c-\gamma)x_2).
$$
The following explores the probability that
$\eta_g(x)$  is located in a neighborhood of the maximum:

\begin{corollary}\label{cor:lm-maxpt} Let $x=(x_1,x_2)\in L$, fix  $r_0\in \R$, and let $n$ be a
positive integer. The probability that the distance between
$\eta_g(x)$ and the point ${\bf a}=(a_1,a_2)$ is between $(n-1)
r_0$ and $n r_0$,  is given in the large $g$ limit by

\beq
 \label{annulus}
 P\Bigl[(n-1)r_0\leq\left|\eta_g(x)-{\bf a}\right|\leq nr_0\Bigr]
&=&\frac{\exp[-\frac{( n^2 + 1 )r_0^2 }{2 \sigma_g^2}]}{4 \pi \sigma_g^4}
\Bigl[e^{\frac{n r_0^2}{\sigma_g^2}}(4 \pi \sigma_g^4+ (n - 1)^2r_0^2\kappa_*^2\log\log g)\nonumber\\
&&~~~~~~~~~~~~~~~~~~~~~~~~~-e^{\frac{r_0^2}{2 \sigma_g^2}}(4\pi \sigma_g^4 + n^2 r_0^2\kappa_*^2\log\log g)\Bigr]
+ O\left(\frac{1}{\log^2 g}\right).
\eeq
\end{corollary}

\vspace{0.3in}

\begin{proof}[Proof:]
 See Appendix.
\end{proof}
\vspace{0.3in}

Notice that the displayed leading terms in equation~(\ref{annulus}) are
independent of the position $x$,  continuous matter density $\kappa_c$, and  external shear $\gamma$.

Corollary~\ref{cor:lm-maxpt}, combined with the lens equation, provides
probabilistic information about the angular position of the
light source.  Indeed, for $n =1$ the corollary gives the
probability of $\eta_g(x)$ lying in a disk of angular radius $r_0$
centered at ${\bf a}$, while for $n \ge 1$ we obtain the
probability of $\eta_g(x)$ lying in an annular region between
radii $(n-1)r_0$ and $ n r_0$ centered at ${\bf a}$. For illustration, we compute
these probabilities in some specific cases:
\[
p_{0.1}(n)=P\Bigl[(n-1)/10\leq\left|\eta_g(x)- {\bf a}\right|\leq n/10\Bigr]
\]
and
$$\tilde{p}_{0.1}(n)=P\Bigl[ 0 \le |\eta_g(x)- {\bf a}|\leq n/10\Bigr]
= \sum_{\ell=1}^n  p_{0.1} (\ell),$$ for $n=1,2,3$ with the
constants $r_0 =0.1$,  $\kappa_* = 0.045$,
 and $g=10^6$.
The probabilities are given in Table I.
\smallskip

\begin{center}
\begin{large}

\begin{tabular}{|c|c|c|c|} \hline
 $n$ &  $1$ & 2& 3 \\ \hline
 ~$p_{0.1} (n)$~ &  ~$0.5595$~ & ~$0.4063$~ & ~$0.0337$~ \\ \hline
 ~$\tilde{p}_{0.1} (n)$~ &  ~$0.5595$~ & ~$0.9658$~ & ~$0.9995$~ \\ \hline
\end{tabular}
\end{large}
\vspace{0.2in}

\mbox{}\\
Table I: Probabilities of a lensing map's value lying in various
annuli (second row) and disks (third row).
\end{center}

The second row of Table I shows that the lensing map value
$\eta_g(x)$ has a $56\%$ probability of lying inside the  disk of
angular radius $r_0 = 0.1$
centered at ${\bf a}$, but a $97\%$ probability of lying within radius $2 r_0$.
 The probability of being in the outermost annulus
$2\, r_0 \leq\left|\eta_g(x)- {\bf a}\right|\leq 3\, r_0$
drops dramatically to $3\%$.

\section{Results of the Random bending angle and fixed points}\label{angle}

We now  use our study of the lensing map to obtain probabilistic
information about the bending angle vector and lensing fixed points.

\subsection{Asymptotic P.D.F. of the Bending Angle}

Recall that the bending angle  vector $\alpha_g(x)$ is given by
$\alpha_g(x) = \nabla \psi_g(x).$ Its asymptotic behavior is given
below:

\begin{corollary}\label{bangle}
Let $x=(x_1,x_2) \in L$. The p.d.f. of  the bending angle vector
is given in the large $g$ limit by

\beq
f_{\alpha_g(x)}(h^*,k^*) &=&\frac{e^{-\frac{(h^*-a_1^*)^2+ (k^*-a_2^*)^2} {2\sigma_g^2}}}
{\Bigl(\sqrt{2\pi}\sigma_g\Bigr)^2}
\Bigr[1-\kappa_*\frac{x_1(h^*-a_1^*)+x_2(k^*-a_2^*)} {\sigma_g^2}\nonumber\\
&& \hspace{1.3in} + \ \frac{\kappa_*^2}{4 \pi} \
\frac{\Bigl((h^*-a_1^*)^2+(k^*-a_2^*)^2-2\sigma_g^2\Bigr)}{\sigma_g^4}\,
\log(\log g)\Bigr] + O\Bigl(\frac{1}{\log^2 g}\Bigr),
\eeq
where
$(h^*,k^*)$ denotes the possible values of the random vector
$\alpha_g(x)$ and $(a_1^*, a_2^*) = ((\kappa_c-\gamma)x_1,
(\kappa_c+\gamma)x_2$, while $\sigma_g$ is defined in
Theorem~\ref{thm-pdf-lm}. In other words, for sufficiently large
$g$, the bending angle is approximately a bivariate normal whose
components $\alpha_{1,g}(x)$ and $\alpha_{2,g}(x)$ are  normal
random variables with respective means
$$
\left[-(\kappa_c + \kappa_*) +\gamma \right]\, x_1 ~~~{\rm and
}~~~ \left[-(\kappa_c + \kappa_*) -\gamma\right]\, x_2,
$$
and variances diverging at leading order as
\[
\frac{\kappa_*^2}{\pi}[\log(B\, g^{1/2}) + \frac{1}{2}\log\log g - \pi x_1^2]~~~{\rm and}~~~
\frac{\kappa_*^2}{\pi}[\log(B\, g^{1/2}) + \frac{1}{2}\log\log g -
\pi x_2^2],
\]
respectively.
\end{corollary}

\vspace{0.3in}

\begin{proof}[Proof:] By definition, we have $\alpha_g(x)=x-\eta_g(x)=\tilde{x}+\N_g(x)$,
where $\tilde{x}=((-\kappa_c+\gamma)x_1,(-\kappa_c-\gamma)x_2)$.
 The corollary then follows from properties of independent random variables,
Theorem~\ref{thm-pdf-lm}, and  some of its corollaries.
\end{proof}
\vspace{0.3in}

When $x={\bf 0}$, the second part of Corollary~\ref{bangle}
recovers, in leading order, earlier results by  Katz, Balbus, and
Paczynski \cite{Katz} and Schneider, Ehlers, and Falco (see
\cite{SchEhlFal92}, p. 325).  Also note that from the discussion
after Theorem~\ref{thm-pdf-lm}, we have that the leading order
term  depends only on the  squared magnitude of the bending
angle due purely to point masses, and the next term breaks this
radial symmetry.

\subsection{ Fixed Points}

Lensed images of a light source usually do not appear at the same
angular position of the light source.  In the case where they do,
they are called {\it fixed points}, which is a notion introduced
in gravitational lensing by Petters and Wicklin \cite{Fixe}.  More
precisely, fixed points of the lensing map $\eta_g$ are points
$x\in L$ such that
\[
\eta_g(x)=x
\]
or equivalently, points where the bending angle vector vanishes.
$\alpha_g(x)={\bf 0}.$ Since $\alpha_g$ is a continuous random
vector, we have $P[\alpha_g(x)= {\bf 0}]=0.$
 Nevertheless, we can study how close $\eta_g(x)$ is to $x$.

\begin{corollary}
 Let $x\in B( {\bf 0},R)\subset L$ and $\epsilon>0$. The probability that $\eta_g(x)$ is
 within $\epsilon$ from $x$ in the light source plane is given by
\[
P\left[\left|\eta_g(x)-x\right|\leq \epsilon\right]=\int_{B({\bf 0},\epsilon)}\,
f_{\alpha_g(x)}(h^*,k^*)\, dh^* dk^*.
\]
\end{corollary}
\vspace{0.3in}

\begin{proof}[Proof:]
This follows directly from Corollary~\ref{bangle} and the above
discussion.
\end{proof}
\vspace{0.3in}

\section{Conclusion}

We presented first steps in the development of a mathematical
theory  of stochastic microlensing that focused on the building
blocks of the theory, namely, the random time delay function and
random lensing map.  We derived exact analytical formulas for the
expectation and  variance of the random time delay function  and
random lensing map about any point. In the large limit of stars,
we found  a simple asymptotic expression for the expectation
and variance of  a normalized time delay function at an
arbitrary point. For the same limit, we highlighted an interesting
link between the leading order term of the expectation and
variance of our  normalized random time delay function and
the first Betti number of its domain. In addition, the asymptotic
p.d.f.s of both the  normalized time delay function and the
scaled lensing map were characterized in the large number of stars
limit: The asymptotic p.d.f. of the former is a shifted gamma
density at leading order, while for the latter the p.d.f. is a
bivariate Gaussian distribution. The p.d.f. of the random scaled
lensing map is also shown to depend on the magnitude of the scaled
bending angle due purely to point masses at leading order and we
illustrated explicitly
 how this radial symmetry fails at the next order. We also derived
 the asymptotic p.d.f. of the random
bending angle vector and gave an estimate of the probability  of a
lens plane point being close to a fixed point. Overall, the paper
determined and illustrated new analytical results about the
microlensing behavior of random time delay functions and random lensing maps
 about an arbitrary point of the lens plane.

In Paper II of this series, we shall explore the
microlensing random shear and expected number of  microimages.

\section{Acknowledgments}
AMT would like to thank Robert Adler, Jonathan Mattingly, and
Andrea Watkins for useful discussions.  AOP acknowledges the
support of NSF Grants DMS-0707003 and AST-0434277-02.

\appendix

\section{Proofs---The Random Time Delay Function}

\noindent {\bf Proposition~1.}
 {\it Let $x\in B({\bf 0},R)$ and $y\in \R^2$ \mbox{\rm(}the light source plane\mbox{\rm)}. Then
\[
E[W_{g,j}(x)]=-\frac{\kappa_*(R-|x|)^2(-1+2\log (R-|x|))}{2g}
-\frac{2\kappa_*}{\pi g }\int_{R-|x|}^{R+|x|} r f(r,|x|)(\log r)
dr,
\]
 and
\beqn
E[W_{g,j}^2(x)]&=&\frac{m\kappa_*(R-|x|)^2 \left[1+2(\log (R-|x|)-1)\log(R-|x|)\right]}{2 g}\\
&&~~~~~~~~~~~~~~~~~~~~~~~+\frac{2m\kappa_*}{\pi
g}\int_{R-|x|}^{R+|x|} r f(r,|x|) (\log^2 r)  dr,
\eeqn
where
$f(r,|x|)={\rm Arccos}\Bigl(\frac{|x|^2+r^2-R^2}{2|x|r}\Bigr)$ for
non-zero $x$, and $f(r,0)=0$.}

\vspace{0.3in}

\begin{proof}[Proof of Proposition~1:] Let $x=(x_1,x_2)\in B({\bf 0},R)\cap L$ and $y\in \R^2$.
Translate the rectangular coordinates $(u,v)$ so that its origin
${\bf 0}$ moves to position $x$. Denote the resulting new
coordinates by $(u',v')$ and its origin by ${\bf 0}^{'}$.  In the
system $(u',v')$, the old origin ${\bf 0}$ now has coordinates
$(u',v')= -x$. Define $\omega_0$ to be the unique principal angle
with $\cos\omega_0=x_1/|x|$, $\sin\omega_0=x_2/|x|$, and let
$\omega \equiv\omega_0+ \pi$. Rotate the rectangular coordinates
$(u',v')$ counterclockwise by angle $\omega$ to obtain new
coordinates $(u^{\prime\prime},v^{\prime\prime})$ with origin
${\bf 0}^{\prime\prime}= {\bf 0}^{\prime}$.  Note that the old
origin $\bf 0$ now lies at position
$(u^{\prime\prime},v^{\prime\prime}) = (|x|,0) \equiv
x^{\prime\prime}$ on the positive $u^{\prime\prime}$-axis.
Finally, let $(\theta, r)$ denote polar coordinates in the frame
$(u^{\prime\prime},v^{\prime\prime})$.  The figure below
illustrate some of these change of variables.

\begin{figure}[htp]
\centering
  \includegraphics[width=3in]{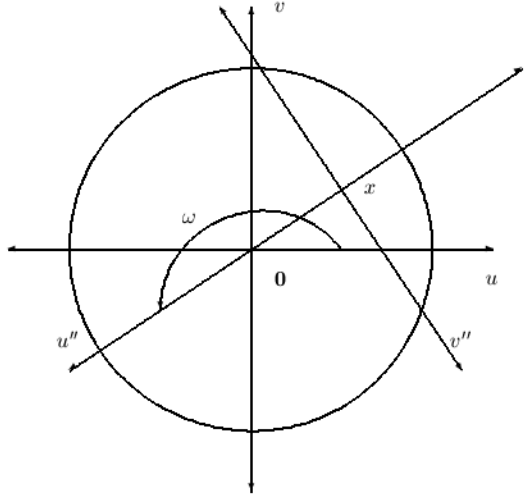}
\caption{Change of variables.}
\label{changevar} 
\end{figure}

We have
$$
E[W_{g,j}(x)]=- mE[\log | x-\xi_j|]
$$
and
\beqn
E[\log\vert x-\xi_j\vert]&=&\frac{1}{2\pi R^2}\int_{B({\bf 0},R)}\log [(u-x_1)^2+(v-x_2)^2]dudv\\
&=&\frac{1}{2\pi R^2}\Bigl[\int_{B({\bf 0}^{'},R-|x|)}\,
\log \left[(u')^2+(v')^2\right]\, du' dv'\\
&&~~~~~~~~~ +\int_{B(-x,R)\backslash B({\bf 0}^{'},R-|x|)}\,
\log \left[(u')^2+(v')^2 \right]\, du' dv' \Bigr].
\eeqn
 For $x={\bf 0}$, the result follows. Suppose $x\neq {\bf 0}$. Then

\beqn
E[\log\vert x-\xi_j\vert] &=&\frac{1}{2\pi R^2}\Bigl[\int_{B({\bf 0}^{\prime\prime},R-|x|)}\,
\log \left[(u^{\prime\prime})^2+(v^{\prime\prime})^2\right]\,
du^{\prime\prime} dv^{\prime\prime}\\
& &~~~~~~~~~ +\int_{B(x^{\prime\prime},R)\backslash B({\bf 0}^{\prime\prime},R-|x|)}\,
\log\left[(u^{\prime\prime})^2+(v^{\prime\prime})^2\right]\,
du^{\prime\prime} dv^{\prime\prime}\Bigr]\\
&=&\frac{1}{\pi R^2}\int_0^{R-|x|}r (\log r)\Bigl(\int_0^{2\pi}d\theta\Bigr)  dr
+\frac{1}{\pi R^2}\int_{R-|x|}^{R+|x|} r(\log r)
\Bigl(\int_{-f(r,|x|)}^{f(r,|x|)}d\theta \Bigr)dr\\
&=&\frac{2}{ R^2}\int_0^{R-|x|}r\log r dr+\frac{2}{\pi R^2}
\int_{R-|x|}^{R+|x|} rf(r,|x|)\log r  dr\\
&=& \frac{(R-|x|)^2(-1+2\log (R-|x|))}{2R^2}+\frac{2}{\pi
R^2}\int_{R-|x|}^{R+|x|} rf(r,|x|) (\log r)  dr,
\eeqn
which gives
the mean. The second moment is obtained similarly.
\end{proof}

\vspace{0.3in}

\noindent {\bf Corollary~2.}
 {\it
Let $x\in B({\bf 0},R)$ and $y\in \R^2$ \mbox{\rm(}the light
source plane\mbox{\rm)}. In the large $g$ limit, we have
\[
E[T_{g,y}^*(x)]= \frac{m}{2}g +O(g^{1/2}\log g)
\]
and
\[
{\rm Var}[T_{g,y}^*(x)]= \frac{m^2}{4}g +O(g^{1/2}\log^2 g).
\]
}

\vspace{0.3in}

\begin{proof}[Proof of Corollary~2:]
Let $d_n(x,y)=\Bigl(\frac{1}{2} | x - y |^2-\frac{\kappa_c}{2} | x |^2 +\frac{\gamma}{2}(x_1^2-x_2^2)\Bigr)/n,\ \ n\in\mathbb{N}$ for arbitrary $x$.

 \beqn
 E[T_{g,y}^*(x)]&=&d_1(x,y)+g E[W_{g,y}(x)]+mg\log R\\
&=&\frac{\kappa_* R^2}{2}\Bigl[\frac{2}{\kappa_*R^2}d_1(x,y)
\ +\ (1-|x|/R)^2-2\frac{|x|^2-2|x|R}{R}\frac{\log R}{R}\\
&&~~~~~~~~-2(1-|x|/R)^2\log (1-|x|/R)-\frac{4}{\pi R^2}
\int_{R-|x|}^{R+|x|} r f(r,|x|) (\log r)  dr\Bigr].
\eeqn
 But
\beqn
\left|\int_{R-|x|}^{R+|x|} r f(r,|x|)(\log r)dr\right|
&\leq&\int_{R-|x|}^{R+|x|} \left|r f(r,|x|)\log r \right| dr\\
&\leq&2\pi \int_{R-|x|}^{R+|x|}r (\log r)  dr~~\mbox{for sufficiently large}~R\\
& \ra & 2 \pi R \log R,
\eeqn
so
$$
\int_{R-|x|}^{R+|x|} r f(r,|x|)(\log r) dr = O(R\log R) = O
(g^{1/2} \log g).
$$
The first result then follows using the relation
$\kappa_*=mg/R^2$. For the second:

\beqn
{\rm Var}[T_{g,y}^*(x)]&=&g{\rm Var}[W_{g,y}(x)]\\
&=&gE[W_{g,y}^2(x)]-g(E[W_{g,y}(x)])^2\\
&=&m\kappa_*\Bigl[\frac{(R-|x|)^2[1+2(\log (R-|x|)-1)\log(R-|x|)]}{2 }+O(R\log^2R)\Bigr]\\
&&-~~~~~g\frac{\kappa_*^2}{g^2}\Bigl[\frac{(R-|x|)^2(-1+2\log (R-|x|))}{2}+O(R\log R)\Bigr]^2\\
&=&m\kappa_*(R-|x|)^2\Bigl[\frac{1}{2}-\frac{(R-|x|)^2}{4R^2}\\
&&~+\Bigl(\log^2(R-|x|)-\log(R-|x|)\Bigr)\Bigl(1-\frac{(R-|x|)^2}{R^2}\Bigr)\Bigr]+O(R\log^2R)\\
&=&\frac{m\kappa_* R^2}{4}+O(R\log^2R).
\eeqn
 This completes the
proof.
\end{proof}

\vspace{0.3in}

\noindent {\bf Proposition~3.}
{\it
For $x\in B({\bf 0},R)$, the p.d.f. of $W_{g,j}(x)$
 is given by:

\beqn
f_{W_{g,j}(x)}(h)&=&\left\{
\begin{array}{lr}
\frac{2}{mR^2}\exp[-\frac{2h}{m}],&~~~~~-m\log (R-|x|)<h\\
A_{R,x}^{'}(h),&~~~~-m\log (R+|x|)<h< -m\log (R-|x|)\\
0,&~~~~h< -m\log (R+|x|),
\end{array}
\right.
\eeqn
 where

\beqn
A_{R,x}(h)  &=&1-\Bigl(r_h\cos^{-1}\theta_{R,1}(h)+R^2\cos^{-1}\theta_{R,2}(h)-B_{R,x}(h)\Bigr)/(\pi R^2),\\
 r_h &=&\exp[2h/m],\\
\theta_{R,1}(h)&=& (|x|^2+r_h^2-R^2)/(2 |x| r_h),\\
\theta_{R,2}(h)&=&(|x|^2+R^2-r_h^2)/(2 |x| R),\\
B_{R,x}(h)&=&\sqrt{(R+r_h-|x|)(d+r_h-R)(d+R-r_h)(d+r_h+R)}/2.
\eeqn
It then follows that the p.d.f. of  $W_{g,j}^*(x)$ is:

\beqn
f_{W_{g,j}^*(x)}(h)=\left\{
\begin{array}{lr}
\frac{2}{m}\exp[-\frac{2h}{m}],&~~~~~-m\log (1-\frac{|x|}{R})<h\\
A_{1,\frac{x}{R}}^{'}(h),&~~~~-m\log (1+\frac{|x|}{R})<h< -m\log (1-\frac{|x|}{R})\\
0,&~~~~h< -m\log (1+\frac{|x|}{R}).
\end{array}
\right.
\eeqn
}

\vspace{0.3in}

\begin{proof}[Proof of Proposition~3:]
Given  $x\in B({\bf 0},R)$, consider the point mass potential
$$W_{g,j}(x) = - m \log|x - \xi_j|.$$
Then

\beq
\label{area}
P(W_{g,j}(x)<h)&=&1-P[|x-\xi|\leq \exp(-h/m)]\nonumber\\
&=&1-\frac{\area{\left(B({\bf 0},R)\cap B(x,R)\right)}}{\pi R^2}\\
&=&\left\{
\begin{array}{lr}
1-\frac{\exp(-2h/m)}{ R^2},&~~~~~0<\exp(-h/m) <R-|x|\\
A_{R,|x|}(h),&~~~~R-|x|< \exp(-h/m) <R+|x|\\
0,&~~~~R+|x|< \exp(-h/m),
\end{array}\right.
\eeq
(see \cite{Eric} for the derivation of the middle entry).
Thus, the function $P(W_{g,j}(x)<h)$ is piecewise smooth in $h$,
and therefore, for almost all $h$ we obtain:

\[
f_{W_{g,j}(x)}(h)=\frac{\partial}{\partial h}P[W_{g,j}(x)<h],
\]
which yields (\ref{pdf-pmpot0}) and (\ref{pdf-pmpot}).
\end{proof}


\vspace{0.3in}

\noindent {\bf Theorem~5.} {\it
 For every $x\in B({\bf 0},R)$,  let
$f_{\frac{T_{g,y}^{*}(x)}{g}}$ be the p.d.f. of
$T_{g,y}^{*}(x)/g$. In the large $g$ limit, we have:

\beqn
f_{\frac{T_{g,y}^{*}(x)}{g}}(h)=f_{\frac{T_{g,y}^{*}({\bf
0})}{g}}(h-d_g(x,y))+O(g^{-1/2}).
\eeqn

}

\vspace{0.3in}

\begin{proof}[Proof of Theorem~5:] Let $h_1=-m\log (1+|x|/R)$,  $h_2=-m\log (1-|x|/R)$, and
$\varphi_{\frac{T_{g,y}^{*}(x)}{g}}$ be the characteristic
function of $\frac{T_{g,y}^{*}(x)}{g}$.  It follows that:

\beqn
\Bigl(e^{-itd_g(x,y)}\varphi_{\frac{T_{g,y}^{*}(x)}{g}}(t)\Bigr)^{1/g}
&=&\frac{2}{m}\int_{h_2}^{\infty}e^{ith/g}e^{-2h/m}dh+\int_{h_1}^{h_2}e^{ith/g}f_{W_{g,j}^{*}(x)}(h)dh\\
&=&(1-\frac{m}{2g}it)^{-1} -  \frac{2}{m}\int_0^{h_2} e^{(it/g-\frac{2}{m})h}dh\\
& & \hspace{1.75in}  +  \int_{h_1}^{h_2}e^{ith/g}f_{W_{g,j}^{*}(x)}(h)dh\\
&=&\varphi_{\frac{T_{g,y}^{*}({\bf 0})}{g}}^{1/g}(t)+O(g^{-3/2}),
\eeqn
where we used the following facts:

\beqn
\int_{h_1}^{h_2}f_{W_{g,j}^{*}(x)}(h)dh&=&P[W_{g,j}^{*}(x)<h_2] = 1-(1-\frac{|x|}{R})^2,\\
-\frac{2}{m}\int_{0}^{h_2}e^{-\frac{2}{m}h}dh&=&(1-\frac{|x|}{R})^2-1.\\
-\frac{2it}{mg}\int_{0}^{h_2}he^{-\frac{2}{m}h}dh&=&\frac{it}{g}\Bigl[h_2(1-P[W_{g,j}^{*}(x)<h_2])
-\frac{m}{2}P[W_{g,j}^{*}(x)<h_2]\Bigr].\\
\frac{it}{g}\int_{h_1}^{h_2}hf_{W_{g,j}^{*}(x)}(h)dh&=&\frac{it}{g}\Bigl[h_2P[W_{g,j}^{*}(x)<h_2]
+O(g^{-1/2})\Bigr],\\
h_2-\frac{m}{2}P[W_{g,j}^{*}(x)<h_2]&=&O(g^{-1}).
\eeqn

 Therefore,

\beqn
e^{-itd_g(x,y)}\varphi_{\frac{T_{g,y}^{*}(x)}{g}}(t)
&=&\varphi_{\frac{T_{g,y}^{*}({\bf 0})}{g}}(t)\left[1+O(g^{-3/2})\right]^g\\
&=&\varphi_{\frac{T_{g,y}^{*}({\bf 0})}{g}}(t)\left[1+O(g^{-1/2})\right]
\eeqn
where the expansion series (sums of terms of order less than $g^{1/2}$)  is the sums of terms of the form $g^q\times t^n$ with $q\in \mathbb{Q}\cap (-\infty,-1/2)$ and $n\in\mathbb{N}$.

By
Proposition~\ref{expo}, the distribution function of
$T_{g,y}^{*}(x)$ is absolutely continuous. Therefore, using  Inverse  Fourier transform, we obtain

\beq
\label{pdf-TDstar0}
f_{\frac{T_{g,y}^{*}(x)}{g}}(h)=f_{\frac{T_{g,y}^{*}({\bf
0})}{g}}(h-d_g(x,y))+O(g^{-1/2}).
\eeq
We can now use the following fact to complete the proof:
\beqn
f_{cX}(h)&=&\frac{1}{c}f_X\left(\frac{h}{c}\right),
\eeqn
where $c>0$ and $X$ is a real-valued random variable.
\end{proof}


\section{Proofs---The Random Lensing Map}

\noindent {\bf Proposition~6.}  {\it For $x=(x_1,x_2)\in L\cap B({\bf 0},R)$, the expectations of the components of
the lensing map are
\[
E[\eta_{1,g}(x)]= \left[1- (\kappa_c + \kappa_*) +\gamma\right]\, x_1
~~~~~~~{\rm and}~~~~~~
E[\eta_{2,g}(x)]= \left[1- (\kappa_c +\kappa_*) -\gamma\right]\, x_2 .
\]
Both variables have infinite variance.
}

\vspace{0.3in}

\begin{proof}[Proof of Proposition~6:] Let $x=(x_1,x_2)\in B({\bf 0},R)\cap L$ and $y\in \R^2$.
Let $f(r,|x|)$ be defined as in the statement of
Proposition~\ref{propp} and employ the coordinates $(u',v')$ and
$(u^{\prime\prime},v^{\prime\prime})$ defined at the start of the
proof of that proposition.

\beqn
E[\eta_{1,g}(x)]&=&(1-\kappa_c +\gamma)x_1+\frac{gm}{\pi R^2}
\int_{B({\bf 0},R)}\frac{(u-x_1)}{(u-x_1)^2+(v-x_2)^2} du dv \\
&=&(1-\kappa_c+\gamma)x_1+\frac{\kappa_*}{\pi}\int_{B({\bf
0},R)\backslash B({\bf 0}^{'},R-|x|)}\, \frac{u'}{(u')^2+(v')^2}\,
du' dv'.
\eeqn
If $x={\bf 0}$, the result follows. For $x \neq {\bf 0}$, we have:
 \beqn
\int_{B({\bf 0},R)\backslash B({\bf 0}^{'},R-|x|)}\,\frac{u'}{(u')^2+(v')^2}\, du' dv'
&=& \int_{B({\bf 0},R)\backslash B({\bf 0}^{\prime\prime},R-|x|)}\,
\frac{u^{\prime\prime} \cos \omega + v^{\prime\prime}  \sin\omega}
{(u^{\prime\prime})^2 - (v^{\prime\prime})^2}\, du^{\prime\prime} dv^{\prime\prime}\\
&=& \int_{R-|x|}^{R+|x|}\int_{-f(r,|x|)}^{f(r,|x|)}
(\cos\theta\cos \omega -  \sin\theta\sin\omega) d\theta dr\\
&=& 2(\cos \omega)\,\int_{R-|x|}^{R+|x|} \sin[f(r,|x|)] dr\\
&=& 2(\cos\omega)\, \int_{R-|x|}^{R+|x|}\sqrt{1-\Bigl(\frac{|x|^2+r^2-R^2}{2r|x|}\Bigr)^2}dr\\
&=&\pi |x|\cos\omega\\
& = & - \pi x_1.
\eeqn
Therefore, $E[\eta_{1,g}(x)] = (1-\kappa_c
-\kappa_* +\gamma)x_1.$ Similarly, $ E[\eta_{2,g}(x)]=(1-\kappa_c
-\kappa_* -\gamma)x_2. $

Next, we establish only the divergence of the variance of
$\eta_{1,g}(x)$ since the proof for the case of $\eta_{2,g}(x)$ is
similar:
\beqn
{\rm Var}[\eta_{1,g}(x)]&=&\sum_{j=1}^g{\rm Var}
\bigl[\frac{ m(U_j - x_1)} {  (U_j - x_1)^2 + (V_j - x_2)^2 }\bigr]\\
&\geq&\frac{gm^2}{\pi R^2}\int_{B({\bf 0},R-|x|)}\frac{u^2}{(u^2+v^2)^2}du dv-
\frac{gm^2}{\pi^2 R^4}(\pi x_1)^2\\
&=&\frac{m\kappa_*}{\pi}\lim_{\varepsilon\ra 0}\int_{\varepsilon}^{ R-|x|}r\int_0^{2\pi} \frac{\sin^2\theta}{r^2} d\theta dr-\frac{m\kappa_*x_1^2}{ R^2}\\
&=&\infty.
\eeqn
\end{proof}

\vspace{0.3in}

\noindent {\bf Theorem~7.}  {\it
Let $x=(x_1,x_2) \in L$ be fixed. The p.d.f. of
the scaled lensing map $\frac{\eta_g(x)}{\sqrt{\log g}}$ is given
in the large $g$ limit by:

\beqn
f_{\frac{\eta_g(x)}{\sqrt{\log g}}}(\tldh, \tldk)
&=&\frac{e^{-\frac{(\tldh-\tilde{a}_1)^2+ (\tldk-\tilde{a}_2)^2}
{2 \tilde{\sigma}_g^2}}}{(\sqrt{2\pi}\tilde{\sigma}_g^2)}
\left[ 1 \ - \ \frac{  \kappa_*}{\sqrt{\log g}}\
\frac{x_1(\tldh-\tilde{a}_1)+x_2(\tldk-\tilde{a}_2)}{\tilde{\sigma}_g^2}\right.
~~~~~~~~~~~~~~~\nonumber\\
&&\left. ~~~~~~~~~~~~~~~+ \frac{\kappa_*^2}{4 \pi} \
\frac{((\tldh-\tilde{a}_1)^2+(\tldk-\tilde{a}_2)^2)-
2\tilde{\sigma}_g^2}{\tilde{\sigma}_g^4}\ \frac{\log(\log g)}{\log g}\right]
+ O\Bigl(\frac{1}{\log g}\Bigr),
\eeqn
 with the constants as previously defined.
The  p.d.f.s of the components of the lensing map are:
\[
f_{\frac{\eta_{i,g}(x)}{\sqrt{\log g}}}(\tilde{c})
=\frac{e^{-\frac{(\tilde{c}-\tilde{a}_i)^2}{2\tilde{\sigma}_g^2}}}{\sqrt{2\pi}\,\tilde{\sigma}_g}
\Bigl[1-\frac{\kappa_*}{\sqrt{\log g}}\frac{x_i (\tilde{c}-\tilde{a}_i)}{\tilde{\sigma}_g^2}+
\frac{\kappa_*^2}{4 \pi}\frac{(\tilde{c}-\tilde{a}_i)^2-\tilde{\sigma}_g^2}{\tilde{\sigma}_g^4}
\,\frac{\log(\log g)}{\log g}\Bigr]+ O\Bigl(\frac{1}{\log g}\Bigr)
\]
for $i=1,2.$

}

\vspace{0.3in}

\begin{proof}[Proof of Theorem~7:]
The scaled lensing map can be written as:
\beq
\label{eq:scaled-eta}
\frac{\eta_{g}(x)}{\sqrt{\log g}}= \frac{\N_g(x)}{\sqrt{\log g}} +
\left(\frac{(1-\kappa_c+\gamma)x_1}{\sqrt{\log
g}},\frac{(1-\kappa_c-\gamma)x_2}{\sqrt{\log g}}\right),
\eeq
where $\frac{\N_g(x)}{\sqrt{\log g}}
=\left(\frac{\N_{1,g}(x)}{\sqrt{\log
g}},\frac{\N_{2,g}(x))}{\sqrt{\log g}}\right)$ with
\[
\N_{1,g}(x)  =  \sum_{j=1}^g  \frac{m(U_j -  x_1)}{R_j^2(x)},
\qquad
 \N_{2,g}(x) =  \sum_{j=1}^g  \frac{m(V_j - x_2)}{R_j^2(x)}.
\]
The possible values of the random vector
$\frac{\eta_{g}(x)}{\sqrt{\log g}}$ will be written
$(\tldh,\tldk)$, while those of $\frac{\N_g (x)}{\sqrt{\log g}}$
will be denoted by $(\tlddh,\tlddk)$.

We begin by determining the asymptotic p.d.f. of
$\frac{\N_g(x)}{\sqrt{\log g}}$. Consider an arbitrary
$x=(x_1,x_2) \in B({\bf 0},R)$. For the random vectors
$(U_i,V_i)$, set $u_x = u - x_1$, $v_x = v-x_2$, and ${\mathfrak
r}_x^2 = u_{x_1}^2 + v_{x_2}^2$. The joint characteristic function
$\varphi_{\frac{\N_g(x)}{\sqrt{\log g}}}(t_1,t_2)$ of
$\frac{\N_g(x)}{\sqrt{\log g}}$ satisfies

\begin{small}
\beq
\label{lastint11}
\left( \varphi_{\frac{\N_g(x)}{\sqrt{\log g}}}(t_1, t_2) \right)^{1/g}
&=& \frac{1}{\pi R^2}  \int_{B({\bf 0},R)}
\exp \Bigl[ \frac{im}{\sqrt{\log g}} \left(\frac{t_1 u_x+ t_2 v_x}{{\mathfrak r_x}^2}\right)
   \Bigr]   \, du dv  \nonumber\\
&=&\frac{2}{ R^2}   \int_0^{R - |x|} r J_0 \Bigl(  \frac{m\sqrt{t_1^2 + t_2^2}}
{ \sqrt{\log g}}r^{-1} \Bigr) \, dr\nonumber\\
&&~~~~~~~~~~~~~~~~~~+   \frac{1}{\pi R^2} \int_{B({\bf 0},R)\backslash B(x,R-|x|)}
\exp \Bigl[ \frac{im}{\sqrt{\log g}} \left(  \frac{t_1 u_x+ t_2 v_x}{{\mathfrak r_x}^2} \right)
   \Bigr]   \, du dv,
\eeq
\end{small}
 where $J_0$ is the zeroth Bessel function.

Adding and subtracting $\frac{2}{R^2}\int_0^{R-|x|} r dr$ to the
first term in
 (\ref{lastint11}) yields:

\beq
\label{lastint122}
\frac{2}{ R^2} \int_0^{R-|x|}  r J_0 \Bigl(\frac{m \sqrt{t_1^2 +t_2^2}}
{ \sqrt{\log g}}r^{-1}  \Bigr)  \, dr
&=&1-\frac{2|x|}{R}+\frac{|x|^2}{R^2}-\frac{2}{ R^2}
\int_0^{R-|x|}  r\left[1- J_0 \Bigl( \frac{m \sqrt{t_1^2 + t_2^2}}{r\sqrt{\log g}} \Bigr) \right] \,  dr\nonumber\\
&=&1-\frac{2|x|}{R}+\frac{|x|^2}{R^2}-\frac{\kappa_*^2(t_1^2+t_2^2)}{4\pi}
\frac{\log (B^2 g)+\log\log g}{g\log g} \nonumber\\
&&~~~~~~~~~~~~~+O(g^{-1}(\log g)^{-1}),
\eeq
 where
 $B=\frac{2e^{1-\gamma_e}}{\kappa_*/\sqrt{\pi}}$  and  we used $\kappa_*=\pi m$.

For  the second integral in (\ref{lastint11}), we get:

\beq
\label{lastint222}
\lefteqn{\frac{1}{\pi R^2} \int_{B({\bf 0},R)\backslash B(x,R-|x|)}
\exp \Bigl[ \frac{im}{\sqrt{\log g}}
\left(  \frac{t_1 u_x+ t_2 v_x}{{\mathfrak r_x}^2} \right) \Bigr] \, du dv}~~~~~~~~~\nonumber\\
~~~~~~~~~~~ &=&  \frac{1}{\pi R^2} \int_{B({\bf 0},R) \backslash B(x,R-|x|)}
\left[1+\frac{im}{\sqrt{\log g}} \left(  \frac{t_1 u_x+ t_2 v_x}{{\mathfrak r_x}^2} \right)\right] du dv  \,+O(g^{-3/2}(\log g)^{-1})\nonumber\\
&=&\frac{2|x|}{R}-\frac{|x|^2}{R^2}+\frac{i|x|\kappa_*}{g\sqrt{\log g}}E(t_1,t_2)+
 O(g^{-3/2}(\log g)^{-1})
\eeq
where $E(t_1,t_2)=t_1\cos\omega +t_2\sin\omega$ with $\omega$
as defined in the proof Proposition \ref{meaneta} and
$\kappa_*=\pi m$.

Combining (\ref{lastint122}) and (\ref{lastint222}), equation~(\ref{lastint11}) becomes,

\beqn
\left(\varphi_{\frac{\N_g(x)}{\sqrt{\log g}}}(t_1, t_2)\right)^{1/g}
&=& 1-\frac{\kappa_*^2(t_1^2+t_2^2)}{4\pi}\frac{\log (B^2 g)}{g\log g}
+\frac{i |x|\kappa_*}{g\sqrt{\log g}}E(t_1,t_2)
 -\frac{\kappa_*^2(t_1^2+t_2^2)}{4\pi}\frac{\log\log g}{ g\log g} + \
 O(g^{-1}(\log g)^{-1}).
\eeqn
Therefore,

\beq
\varphi_{\frac{\N_g(x)}{\sqrt{\log g}}}(t_1, t_2)
&=&\Bigl[1-\frac{1}{g}\Bigl(\frac{\kappa_*^2(t_1^2+t_2^2)}{4\pi}
\frac{\log(B^2 g)}{\log g}\Bigr)\Bigr]^g \Bigl[1 + \frac{i|x|\kappa_*}{g\sqrt{\log g}} E(t_1,t_2) - \frac{\kappa_*^2(t_1^2+t_2^2)}{4\pi}\frac{\log\log g}{ g\log g} \nonumber\\
&&~~~~~~~~~~~~~~~~~~~~~~~~~~~~~~~~~~~~~~~~~~~~~~~~~~~~~~~~~~+
O\left(g^{-1}(\log g)^{-1}\right)\Bigr]^g\nonumber\\
&=&\Bigl[1 - \frac{1}{g}\Bigl(\frac{\kappa_*^2(t_1^2+t_2^2)}{4\pi}
\frac{\log (B^2 g)}{\log g}\Bigr)\Bigr]^g \Bigl[1+\frac{i |x|\kappa_*}{\sqrt{\log g}} E(t_1,t_2)-\frac{\kappa_*^2(t_1^2+t_2^2)}{4\pi}\frac{\log\log g}{ \log g} \nonumber\\
&&~~~~~~~~~~~~~~~~~~~~~~~~~~~~~~~~~~~~~~~~~~~~~~~~~~~~~~~~~~+
O\left((\log g)^{-1}\right)\Bigr]\nonumber\\
&=& \exp\Bigl[ - \frac{(t_1^2+t_2^2)\sigma_g^2}{2\log g}\Bigr]\Bigl[1+
\frac{i  |x|\kappa_*}{\sqrt{\log g}}E(t_1,t_2)-\frac{\kappa_*^2(t_1^2+t_2^2)}{4\pi }
\frac{\log (\log g)}{\log g}\Bigr] +  O\left(\frac{1}{\log g}\right). \label{eq:proof-thm-apdf-2}
\eeq

The term  remainder term in equation~(\ref{eq:proof-thm-apdf-2}) is a sum of term of the form $p(g)\times q(t_1,t_2)$ where $p(g)$ has order  $1/\log g$  or less and $q(t_1,t_2)$ is independent of $g$ and  integrable. We can, therefore,  take the Inverse Fourier transform (I.F.T.) of
(\ref{eq:proof-thm-apdf-2}) to obtain the joint p.d.f of
$\frac{\N_g(x)}{\sqrt{\log g}}$. Since the I.F.T. of the L.H.S. is
well defined due to $\eta_g (x)$ having an absolutely continuous
cumulative distribution function, the I.F.T. of the R.H.S. of
equation~(\ref{eq:proof-thm-apdf-2}) carries through, and can be
computed term by term. The I.F.T. of the first term in
(\ref{eq:proof-thm-apdf-2}) is a standard integral, and gives

\beqn
\frac{1}{(2\pi)^2}\int_{\R^2}\exp\left[-i(t_1\tlddh+t_2\tlddk)\right]\,
\exp\left[-\frac{(t_1^2+t_2^2) \sigma_g^2}{2\log g}\right]dt_1dt_2
&=&\frac{e^{-\frac{\tlddh^2+\tlddk^2}{2 (\sigma_g^2/\log g)}}}
{2\pi (\sigma_g/\sqrt{\log g}))^2}.
 \eeqn
For the second term in (\ref{eq:proof-thm-apdf-2}), we
obtain

\beqn
\lefteqn{\frac{1}{(2\pi)^2} \int_{\R^2} \exp \left[-i(t_1 \tlddh + t_2 \tlddk)\right]\,
\exp\left[-\frac{(t_1^2+t_2^2) \sigma_g^2}{2\log g}\right]
\frac{i |x|\kappa_*}{\sqrt{\log g}} t_1 \cos \omega dt_1dt_2}~~~~~~~~~~~~~~~~~~~~~~~~~~~~~\\
&=&\frac{i |x|\kappa_*\cos \omega}{(2\pi)^2\sqrt{\log g}}
\int_0^{\infty}\,r\, \int_0^{2\pi}r\cos\theta \exp\left[-\frac{ r^2 \sigma_g^2}{2\log g}\right] \exp\left[-i r\sqrt{h^2+k^2}\cos(\theta + \phi)\right]\,d\theta dr\\
&=&\frac{i |x|\kappa_*\cos \omega}{(2\pi)^2\sqrt{\log g}} \int_0^{\infty}\,r^2\,\exp\left[-\frac{r^2\sigma_g^2}{2\log g}\right]\,
\int_{\phi}^{2\pi + \phi} \sin(\theta +  \frac{\pi}{2}-\phi)
\exp\left[-i r \sqrt{h^2+k^2} \cos \theta \right]\,d\theta dr\\
&=&\frac{ |x|\kappa_*\cos \omega\cos\phi }{2\pi\sqrt{\log g}} \int_0^{\infty}\,r^2\,\exp\left[-\frac{r^2\sigma_g^2}{2\log g}\right]\,
J_1\left(r\sqrt{h^2+k^2}\right)\, dr,\\
\eeqn
where
$\tlddh\cos\nu+\tlddk\sin\nu=\sqrt{\tlddh^2+\tlddk^2}\cos(\nu
+\phi),\,\,\,\forall\,\nu$.

Similarly,

\beqn
\lefteqn{\frac{1}{(2\pi)^2}  \int_{\R^2}\exp\left[-i (t_1 \tlddh +t_2 \tlddk)\right]\,
\exp\left[-\frac{(t_1^2 + t_2^2) \sigma_g^2}{2\log g}\right]
\frac{i |x|\kappa_*}{\sqrt{\log g}} t_1 \cos \omega dt_1 dt_2}~~~~~~~~~~~~~~~~~~~~~~~~~~~~~\\
&=&\frac{- |x|\kappa_*\sin \omega  \sin \phi }{2\pi\sqrt{\log g}}
\int_0^{\infty}\,r^2\,\exp \left[-\frac{r^2\sigma_g^2}{2\log g}\right]\,
 J_1\left(r\sqrt{h^2+k^2}\right)\, dr.
\eeqn
The
sums gives
\beqn
\lefteqn{\frac{1}{(2\pi)^2} \int_{\R^2}\exp\left[-i( t_1 \tlddh + t_2\tlddk)\right]\,
\exp\left[-\frac{(t_1^2 + t_2^2) \sigma_g^2}{2\log g}\right]
\frac{i |x|\kappa_*}{\sqrt{\log g}}(t_1 \cos \omega + t_2\sin \omega)\ dt_1dt_2} ~~~~~~~~~~~~~~~~~~~~~~~~~~~~~\\
&=&\frac{ |x|\kappa_*\cos( \omega + \phi) }{2\pi \sqrt{\log g}}
\int_0^{\infty}\,r^2\,\exp\left[-\frac{r^2 \sigma_g^2}{2\log g}\right]\,
J_1 \left(r \sqrt{\tlddh^2 + \tlddk^2}  \right)\, dr.\\
&=&\frac{ |x|\kappa_*}{2\pi \sqrt{\log g}}\sqrt{\tlddh^2 + \tlddk^2} \cos( \omega+\phi)
\frac{e^{-\frac{\tlddh^2 + \tlddk^2}{2 (\sigma_g^2/\log g)}}} {  (\sigma_g^2/\log g)^2}\\
&=& \frac{ |x|\kappa_*}{2\pi\sqrt{\log g}} \tlddh \cos \omega + \tlddk \sin \omega)
\frac{e^{-\frac{\tlddh^2 + \tlddk^2}{2(\sigma_g^2/\log g)}}}{(\sigma_g^2/\log g)^2}.
\eeqn

The
 I.F.T. of the  third term in (\ref{eq:proof-thm-apdf-2}) yields:
\beqn
\lefteqn{\frac{1}{(2\pi)^2}\int_{\R^2}(t_1^2+t_2^2)\exp\left[-i(t_1\tlddh+t_2\tlddk)\right]\,
\exp\left[-\frac{(t_1^2+t_2^2) \sigma_g^2}{2\log g}\right] dt_1 dt_2}
~~~~~~~~~~~~~~~~~~~~~~~~~~~~~~~~~~~~~~~~~~~~~~~~~~\\
&=&\frac{1}{(2\pi)^2} \int_{0}^{\infty} r^{3} e^{-\frac{r^2 \sigma_g^2}{2\log g}}
\left(\int_0^{2\pi} e^{-ir\sqrt{\tlddh^2+\tlddk^2} \cos(\theta+\phi)}d\theta\right) dr\\
&=&\frac{1}{2\pi} \int_{0}^{\infty}r^{3}e^{-\frac{r^2 \sigma_g^2}{2\log g}}
J_0\left(r\sqrt{\tlddh^2+\tlddk^2}\right) dr\\
&=&-\frac{e^{-\frac{\tlddh^2+\tlddk^2}{2 (\sigma_g^2/\log g)}}}
{2\pi (\sigma_g/\sqrt{\log g}))^6}
\left(\tlddh^2 + \tlddk^2-2(\sigma_g/\sqrt{\log g})^2\right).
 \eeqn
Finally, we combine these results to get:

\beqn
f_{\frac{\N_{1,g}(x)}{\sqrt{\log g}},\frac{\N_{2,g}(x)}{\sqrt{\log g}}}(\tlddh,\tlddk)
&=&\frac{e^{-\frac{\tlddh^2 + \tlddk^2}{2 (\sigma_g^2/\log g)}}}{(\sqrt{2\pi}
(\sigma_g/\sqrt{\log g}))^2}
\left[ 1 \ - \ \frac{  \kappa_*}{\sqrt{\log g}}\
\frac{x_1\tlddh+x_2\tlddk} {(\sigma_g^2/\log g)}\right.~~~~~~~~~~~~~~~\\
&&\left. ~~~~~~~~~~~~~~~+ \frac{\kappa_*^2}{4 \pi} \
\frac{(\tlddh^2+\tlddk^2)- 2 (\sigma_g^2/\log g)}{(\sigma_g^2/\log g)^2}\
\frac{\log(\log g)}{\log g}\right]+ O\Bigl(\frac{1}{\log g}\Bigr).
\eeqn

Returning to $\frac{\eta_g(x)}{\sqrt{\log g}}$ in equation~(\ref{eq:scaled-eta}), we note that

\beqn
  f_{\frac{\eta_{g}(x)}{\sqrt{\log  g}} } (\tldh,\tldk)
&=& f_{\frac{\N_g(x)}{\sqrt{\log g}}} \Bigl(\tldh-\frac{(1-\kappa_c+\gamma)x_1}{\sqrt{\log g}},
\tldk-\frac{(1-\kappa_c-\gamma)x_2}{\sqrt{\log g}}\Bigr).
 \eeqn
The first part of the theorem now follows; the second part is a
consequence of the first by integration.
\end{proof}

\vspace{0.3in}

{\bf Corollary~8.}
{\it
As $g \ra \infty$ with fixed $x \in \R^2$, we have
\[
     \Bigl( \frac{ \eta_{1,g}(x)}{ \sqrt{\log g} }   ,  \frac{\eta_{2,g}(x)}{\sqrt{\log
 g}}   \Bigr)
      \Rightarrow   ( \eta_{1,\infty}, \eta_{2,\infty} ).
\]
Here ``$\Rightarrow$'' denotes distributional convergence and
$(\eta_{1,\infty}, \eta_{2,\infty} )$ is bivariant normal random
vector with independent mean-zero normal random variables as
components, each with variance $\kappa_{*}^2/(2\pi)$.
}
\vspace{0.3in}

\begin{proof}[Proof of Corollary~8:]
This result follows directly from
equation~(\ref{eq:proof-thm-apdf-2}) in the proof of
Theorem~\ref{thm-pdf-lm}.
\end{proof}

\vspace{0.3in}

\noindent {\bf Corollary~11.}  {\it
Let $x=(x_1,x_2)\in L$, fix  $r_0\in \R$, and let $n$ be a
positive integer. The probability that the distance between
$\eta_g(x)$ and the point ${\bf a}=(a_1,a_2)$ is between $(n-1)
r_0$ and $n r_0$,  is given in the large $g$ limit by

\beqn
 P\Bigl[(n-1)r_0\leq\left|\eta_g(x)-{\bf a}\right|\leq nr_0\Bigr]
&=&\frac{\exp[-\frac{(n^2+1)r_0^2}{2\sigma_g^2}]}{4\pi\sigma_g^4}
\Bigl[e^{\frac{nr_0^2}{\sigma_g^2}}(4\pi\sigma_g^4+(n-1)^2r_0^2\kappa_*^2)\nonumber\\
&&~~~~~~~~~~~~~~~~~~~-e^{\frac{r_0^2}{2\sigma_g^2}}(4\pi\sigma_g^4+n^2r_0^2\kappa_*^2)H(g)\Bigr]
+ O\left(\frac{1}{\log^2 g}\right).
\eeqn

}

\vspace{0.3in}

\begin{proof}[Proof of Corollary~11:]
Set $p_{r_0}(n)=P\Bigl[(n-1)r_0\leq\left|\eta_g(x)-{\bf
a}\right|\leq nr_0\Bigr]$.
\beqn
p_{r_0}(n) &=& P\Bigl[\eta_g(x)\in A((a_1,a_2),(n-1)\cdot r_0,n\cdot r_0)\Bigr]\\
&=&\int_{A\left({\bf a},(n-1)r_0,nr_0\right)}\ {\cal F}_{g,x}(h,k)\ dh dk \ +  O\left( \frac{1}{\log^2 g} \right)\\
&=&\int_{(n-1)r_0}^{nr_0}\frac{re^{-\frac{r^2}{2\sigma_g^2}}}{\sigma_g^2}
\Bigl[1 +  \frac{\kappa_*^2}{4 \pi}\ \frac{r^2-2 \sigma_g^2}{ \sigma_g^4} H(g)\Bigr] dr
+  O\left( \frac{1}{\log^2 g} \right)\\
&=&\frac{\exp[-\frac{( n^2 + 1 )r_0^2 }{2 \sigma_g^2}]}{4 \pi \sigma_g^4}
\left[e^{\frac{n r_0^2}{\sigma_g^2}}(4 \pi \sigma_g^4+ (n - 1)^2r_0^2\kappa_*^2\log\log g)
-e^{\frac{r_0^2}{2 \sigma_g^2}}(4\pi \sigma_g^4 + n^2 r_0^2\kappa_*^2\log\log g)\right]
+ O\left(\frac{1}{\log^2 g}\right).
\eeqn

\end{proof}


\begin{thebibliography}{}


\bibitem{Schechter1}
P. L. Schechter, and J. Wambsganss, Astrophysics J., {\bf 580} 685
(2002).




\bibitem{Vietri}

M. Vietri,  and J. P. Ostriker,   Astrophysics J., {\bf 267}, 488
(1983).



\bibitem{Kayser}
R. Kayser, S. Refsdal,  and R. Stabell, Astron. Astrophys., {\bf
166}, 36 (1986).



\bibitem{Pac}
B. Paczynski,  Astrophysics J.., {\bf 301}, 503 (1986).



\bibitem{Deguchi}
S. Deguchi, and W. D. Watson,   Astrophysics J., {\bf 335}, 67
(1988).




\bibitem{Deguchi2}
\underline{~~~~~~~~~~}, Phys. Rev. Letters, {\bf 59}, 2814,
(1987).



\bibitem{Katz}
N. Katz, S. Balbus,  and B. Paczynski, Ajp, {\bf 306}, 2, (1986).



\bibitem{Schneider}
P. Schneider,   Astrophysics J.,  {\bf 319} 9 (1987).




\bibitem{Turner}
E. L. Turner, J. P.  Ostriker,  and J. R. Gott,  Astrophysics J..,
{\bf 284}, 1 (1984).



\bibitem{Vietri1}

M. Vietri,   Astrophysics J., {\bf 293}, 343 (1985).



\bibitem{Rauch}
K. P. Rauch, S. Mao,  J.  Wambsganss, and B. Paczynski,
Astrophysics J., {\bf 386}, 30 (1992).



\bibitem{Wambsganss}
J. Wambsganss,  Astrophysics J.. {\bf 386}, 19, (1992).



\bibitem{wamb}
J. Wambsganss, H. J.  Witt, and P. Schneider, Astron. Astrophys.,
{\bf 258}  591 (1992).



\bibitem{GSW}
J. Granot, P. L. Schrechter, and J.   Wambsganss, Ajp, {\bf 583},
575, (2003).



\bibitem{Keeton}
R. C. Keeton,  and  L. A. Moustakas,  e-print arXiv:0805.0309.



\bibitem{Schechter2}
P. L. Schechter, J.  Wambsganss, and G. F. Lewis,  Astrophysics J.
{\bf 613},  77 (2004).



\bibitem{Tuntsov}
A. V. Tuntsov, G. F.  Lewis,R. A.  Ibata,  and J. P. Kneib,
Mon.Not. R. Astron. Soc. {\bf 000}, 1 15 (2004).



\bibitem{Berry}
M. V. Berry, and  C. Upstill, in {\it Progress in Optics}, E.
Wolf, ed. (North Holland, Amsterdam,  Vol. XVIII, 1980).






\bibitem{Adlerjon}
R. Adler and J. Taylor, {\it Random Fields and Geometry} (Wiley,
London, 1981).







\bibitem{Fixe}
 A. O. Petters, and  F. J. Wicklin,  J. Math. Phys., {\bf 39}, 1011 (1998).



\bibitem{Petters}
A. O. Petters, H. Levine, and J. Wambsganss, {\it Singularity
Theory and Gravitational Lensing} (Birkh\"auser, 2001).



\bibitem{Eric}
E. W. Weistein; ``Circle-Circle Intersection." From MathWorld--A Wolfram Web Resource.\\
 http://mathworld.wolfram.com/Circle-CircleIntersection.html






\bibitem{SchEhlFal92}
P. Schneider, J. Ehlers, and E. E. Falco, {\it Gravitational
Lenses} (Springer, Berlin, 1992).




\end{thebibliography}
\end{document}